\documentclass[aps,pra,preprint,tightenlines,showpacs,groupedaddress]{revtex4}
\usepackage{graphicx}
\usepackage{amsmath}
\usepackage{longtable}

\begin{document}

\title{Convergence of CI single center calculations of positron-atom interactions }
\author{J.Mitroy}
\email{jxm107@rsphysse.anu.edu.au}
\affiliation{Faculty of Technology, Charles Darwin University, Darwin NT 0909, Australia}
\author{M.W.J.Bromley}
\email{mbromley@physics.sdsu.edu}
\affiliation{Department of Physics, San Diego State University, San Diego CA 92182, USA}

\date{\today}

\begin{abstract}

The Configuration Interaction (CI) method using orbitals centered on the 
nucleus has recently been applied to calculate the interactions of 
positrons interacting with atoms.  Computational investigations of the 
convergence properties of binding energy, phase shift and annihilation
rate with respect to the maximum angular momentum of the orbital basis 
for the $e^+$Cu and PsH bound states, and the $e^+$-H 
scattering system were completed.  The annihilation rates converge very 
slowly with angular momentum, and moreover the convergence with radial 
basis dimension appears to be slower for high angular momentum.  A 
number of methods of completing the partial wave sum are compared, an 
approach based on a $\Delta X_J =  
a(J + {\scriptstyle \frac{1}{2}})^{-n} +  b(J + {\scriptstyle \frac{1}{2}})^{-(n+1)}$ 
form (with $n = 4$ for phase shift (or energy) and $n = 2$ for the annihilation 
rate)  seems to be preferred on considerations of utility and underlying 
physical justification.

\end{abstract}


\maketitle

\section{Introduction}

In the last few years there have been a number of calculations 
of positron-binding to atoms 
\cite{mitroy99c,dzuba99,dzuba00a,bromley00a,bromley02a,bromley02b,bromley02d,bromley02e,saito03a,saito03c,saito05a} 
and positron-scattering from atoms and ions 
\cite{bromley03a,novikov04a,gribakin04b}
using orthodox CI type methods with large basis sets of 
single center orbitals.   Besides the calculations on atoms, 
there have some attempts to calculate
the properties of positrons bound to molecules with modified 
versions of standard quantum chemistry methods 
\cite{tachikawa03a,strasburger04a,buenker05a}.      

One feature common to all these CI type calculations is the  
slow convergence of the binding energy, and the even slower 
convergence of the annihilation rate. The 
attractive electron-positron interaction leads to the formation of a 
Ps cluster (i.e. something akin to a positronium atom) in the outer 
valence region of the atom \cite{ryzhikh98e,dzuba99,mitroy02b,saito03a}.   
The accurate representation of a Ps cluster using only single particle 
orbitals centered on the nucleus requires the inclusion of orbitals 
with much higher angular momenta than a roughly equivalent electron-only 
calculation \cite{strasburger95,schrader98,mitroy99c,dzuba99}.  For example, 
the largest CI calculations on PsH and the group II positronic atoms have 
typically involved single particle basis sets with 8 radial functions per 
angular momenta, $\ell$, and inclusion of angular momenta up to 
$\ell_{\rm max} = 10$ \cite{bromley02a,bromley02b,saito03a}.  Even with such 
large basis sets, between 5-60$\%$ of the binding energy and some 
30-80$\%$ of the annihilation rate were obtained by extrapolating 
from the $\ell_{\rm max} = 10$ to the $\ell_{\rm max} = \infty$ limit. 
 
Since our initial CI calculations on group II and IIB atoms
\cite{bromley02a,bromley02b,bromley02d}, advances in computer hardware mean larger 
dimension CI calculations are possible.  In addition, program improvements 
have removed the chief memory bottleneck that previously constrained the 
size of the calculation.  As a result, it is now appropriate to revisit 
these earlier calculation to obtain improved estimates of the positron 
binding energies and other expectation values.  However, as the calculations
are increased in size, it has become apparent that the issue of slow
convergence of the physical observables with the angular momenta 
of the basis orbitals is the central technical issue in any calculation.  

Whilst it is desirable to minimize the amount of mechanical detail in 
any discussion (so as not to distract from the physics), the ability to
draw reliable conclusions from any calculation depends crucially
on the treatment of the higher partial waves.  For example, the
treatment of the higher partial waves in separate calculations by Saito 
\cite{saito05a} has already been shown to be flawed \cite{mitroy05i} 
while the present work exposes the defects in the methods of Gribakin 
and Ludlow \cite{gribakin04b}.  The present work, 
therefore, is solely devoted to an in-depth examination of the    
convergence properties of mixed electron-positron calculations.  
 
In our previous works, 
\cite{bromley02a,bromley02b,bromley02d,bromley03a,novikov04a}, a
relatively simple solution to this problem was adopted.  In effect, 
it was assumed that the successive increments to any observable scaled
as an inverse power series in $\ell_{\rm max}$.  This approach does 
have limitations as do the approaches adopted by other groups 
\cite{saito03c,saito05a,gribakin04b,mitroy05i}.   
In the present article, we examine the convergence properties of 
positron binding calculations upon PsH and $e^+$Cu, and a positron 
scattering calculation upon the $e^+$-H system with respect to 
angular momentum and the dimension of the radial basis sets.  The 
limitations of existing calculations are exhibited, and some improved
prescriptions for estimating the variational limit are introduced 
and tested.     

\section{Existing methods of performing the angular momentum extrapolation}

\subsection{The nature of the problem}  

The positron-atom wave function is written 
as a linear combination of states 
created by multiplying atomic states to single particle positron 
states with the usual Clebsch-Gordan coupling coefficients ; 
\begin{eqnarray}
|\Psi;LS \rangle & = & \sum_{i,j} c_{i,j} \ 
\langle L_i M_i \ell_j m_j|L M_L \rangle  
\langle S_i M_{S_i} {\scriptstyle \frac{1}{2}} \mu_j|S M_S \rangle \nonumber \\  
& \times & \ \Phi_i(Atom;L_iS_i) \phi_j({\bf r}_0) \ .   
\label{wvfn} 
\end{eqnarray}
In the case of a single electron system, e.g. H, $\Phi_i(Atom;L_i S_i)$ is  
just a single electron wave function, i.e. an orbital.  For a di-valent
system, $\Phi_i(Atom;L_i S_i)$ is an antisymmetric product of two single 
electron orbitals coupled to have good $L_i$ and $S_i$ quantum numbers.  
The function $\phi_j({\bf r}_0)$ is a single positron orbital.  The 
single particle orbitals are written as a product of a radial function 
and a spherical harmonic:
\begin{equation}
\phi({\bf r})  =  P(r) Y_{\ell m}({\hat {\bf r}}) \ .         
\label{orbital} 
\end{equation}
The radial wave functions are a linear combination of Slater Type
Orbitals (STO) \cite{mitroy99f} and Laguerre Type Orbitals (LTOs).   
Most of the time the radial functions are LTOs, the exceptions 
occurring for single electron states with angular momenta equal 
to those of any occupied core orbitals.  Since the Hartree-Fock 
core orbitals are written as a single combination of STOs, some
of the active electron basis is written as linear combinations of
STOs before the switch to a LTO basis is made.  The LTO basis 
\cite{bromley02a,bromley02b} has the property that the basis can 
be expanded toward completeness without having any linear 
independence problems.  
 
The present discussion is specific to positronic systems with a 
total orbital angular momentum of zero.  It is straight-forward 
to generalize the discussion to states with $L > 0$, but this
just adds additional algebraic complexities without altering any
of the general conclusions.

For a one electron system, the basis can be characterized by the index
$J$, the maximum orbital angular momentum of any single electron
or single positron orbital included in the expansion of the wave function.

For two electron systems, the $L = 0$ configurations are generated 
by letting the two electrons and positron populate the single particle 
orbitals subject to two selection rules,
\begin{eqnarray}
\max(\ell_0,\ell_1,\ell_2) & \le & J \ ,  \\ 
\min(\ell_1,\ell_2) & \le & L_{\rm int} \ .   
\label{CIselect} 
\end{eqnarray}
In these rules $\ell_0$ is the positron orbital angular momentum, 
while $\ell_1$ and $\ell_2$ are the angular momenta of the 
electrons.   The maximum angular momentum of any electron or 
positron orbital included in the CI expansion is $J$.  The 
other parameter, $L_{\rm int}$ is used to eliminate configurations 
involving the simultaneous excitation of both electrons into high 
$\ell$ states.  Double excitations of the two electrons into excited
orbitals are important for taking electron-electron correlations into
account, but the electron-electron correlations converge a lot more
quickly with $L_{\rm int}$ than electron-positron correlations do with
$J$.  Calculations of the positronic bound states of the 
group II atoms and PsH \cite{bromley02a,bromley02b} showed that the 
annihilation rate changed by less than 1$\%$ when $L_{\rm int}$ was varied 
from 1 to 3.   The present set of calculations upon PsH were performed 
with $L_{\rm int} = 4$.  Further details about the methods used to perform 
the calculations can be found elsewhere \cite{bromley02a,bromley02b}.  

Various expectation values are computed to provide information 
about the structure of these systems.  All observable quantities 
can be defined symbolically as  
\begin{equation}
\langle X \rangle^{J} = \sum_{L=0}^{J} \Delta X^{L} \ ,  
\label{XJ1}
\end{equation}
where $\Delta X^{J}$ is the increment to the observable that occurs
when the maximum orbital angular momentum is increased from 
$J\!- \! 1$ to $J$, e.g.     
\begin{equation}
\Delta X^{J} = \langle X \rangle^{J} - \langle X \rangle^{J-1} \ .  
\label{XJ3}
\end{equation}
Hence, one can write formally 
\begin{equation}
\langle X \rangle^{\infty} = \langle X \rangle^{J}  + \sum_{L=J+1}^{\infty} \Delta X^{L} \ .  
\label{XJ2}
\end{equation}
The first term on the right hand side will be determined by explicit computation
while the second term must be estimated.  The problem confronting all single 
center calculations is that most expectation values, $\langle X \rangle^{J}$ 
converges  relatively slowly with $J$ and so the contribution of the second 
term can be significant.  A sensible working strategy is to make $J$ 
as large as possible while simultaneously trying to use the best possible 
approximation to mop-up the rest of the partial wave sum.      

\subsection{Existing extrapolation techniques and their limitations}

One of the first groups to confront this issue and attempt a solution 
was the York University group of McEachran and Stauffer.
They performed a series of polarized orbital calculations of 
positron scattering from rare gases 
\cite{mceachran78a,mceachran78b,mceachran79,mceachran80}.  The decrease in  
energy when the target atom relaxed in the field of a fixed positron was 
used to determine the polarization potential as a function of 
the distance from the nucleus.  The polarized orbital method implicitly 
includes the influence of virtual Ps formation (within an adiabatic 
approximation), and this means that slow convergence can be expected.  
McEachran and Stauffer found that the scattering observables, namely 
the phase shift, and the $Z_{\rm eff}$ annihilation parameter, converged 
slowly with $J$, the largest angular momentum of the polarized electron
orbital set used to represent the adjustment of the atomic charge cloud 
in the field of the positron.  They took this into consideration by 
assuming their polarization potential scaled as $J^{-p}$ and the
polarized orbital scaled as $J^{-q}$ at large $J$. They found 
$p \approx 3.8$ and $q \approx 1.8$ for the rare gases at 
$J \approx 12$.   

The recent CI-type calculations of Mitroy and collaborators also used an 
inverse power relation of $J^{-p}$, to complete the partial wave sum 
\cite{bromley02a,bromley02b,bromley03a}.   In this case, the observables, 
$\varepsilon^J$, $\delta^J$, $\Gamma^J$ and $Z^J_{\rm eff}$ were 
extrapolated.  This contrasts with the polarized orbital calculations 
where the polarization potential and polarized orbital were extrapolated.  
The value of $p$ was given by   
\begin{equation}
p =   \ln \left(  \frac {\Delta X^{J-1}}{\Delta X^J} \right) \biggl/ 
      \ln \left( \frac{J-1}{J} \right) \ ,  
\label{pfraction1}
\end{equation}
while the constant factor is  
\begin{equation}
A_X =   \Delta X^J J^{p}  \ .   
\label{Avalue}
\end{equation}
The correction factor was then evaluated by doing the sum 
$\sum_{L=J+1}^{\infty} A_X/L^p$ explicitly with an upper     
limit in the thousands.   
    
Gribakin and Ludlow \cite{gribakin02a} applied perturbation theory
and the ideas of Schwartz \cite{schwartz62a,schwartz62b} to determine the 
asymptotic behavior of the partial wave increments to the binding energies, 
phase shifts and annihilation rates of positron-atom systems.  This work 
is largely derived from previous work on the partial wave expansion of two 
electron atoms \cite{hill85a,kutzelnigg92a,schmidt93a,ottschofski97a}.    
They determined that the binding energy $E$, annihilation rate $\Gamma$, 
phase shift $\delta$, and collisional annihilation parameter $Z_{\rm eff}$  
obey 
\begin{eqnarray} 
\Delta E^{J}  =  \langle E \rangle^J - \langle E \rangle^{J-1} &\sim & \frac{B_E}{(J+{\scriptstyle \frac{1}{2}})^4}  \label{pE} \\   
\Delta \Gamma^J = \langle \Gamma \rangle^J - \langle \Gamma \rangle^{J-1} & \sim & \frac{B_{\Gamma}}{(J+{\scriptstyle \frac{1}{2}})^2} \label{pGamma} \\   
\Delta \delta^J  = \langle \delta \rangle^J - \langle \delta \rangle^{J-1} & \sim & \frac{B_{\delta}}{(J+{\scriptstyle \frac{1}{2}})^4} \label{pdelta}  \\   
\Delta Z^J_{\rm eff} = \langle Z _{\rm eff} \rangle^J - \langle Z_{\rm eff} \rangle^{J-1} & \sim & \frac{B_Z}{(J+{\scriptstyle \frac{1}{2}})^2} \label{pZ}     
\end{eqnarray} 
These expressions are merely the leading order terms of a series 
of the form  
\begin{eqnarray} 
\Delta X^J = \frac{B_X}{(J+{\scriptstyle \frac{1}{2}})^n} 
   + \frac{C_X}{(J+{\scriptstyle \frac{1}{2}})^{n+1}} + \ldots 
\label{Zseries}     
\end{eqnarray} 
To perform the actual extrapolation during a calculation of positron-hydrogen
scattering, Gribakin and Ludlow \cite{gribakin04b} did a fit to calculated 
values at $J = 9$ and $J = 10$ with the formulae 
\begin{eqnarray}
\delta =  \langle \delta \rangle^{\infty} &=&  \langle \delta \rangle^J + \sum_{L=J+1}^{\infty}
\frac{B_{\delta}}{(L+{\scriptstyle \frac{1}{2}})^4 } \nonumber \\  
&\approx&  \langle \delta \rangle^J + \frac{B_{\delta}}{3(J+{\scriptstyle \frac{1}{2}})^3 }  \ 
\label{Glebfit1} \ , \\  
Z_{\rm eff}  = \langle Z_{\rm eff} \rangle^{\infty}  
&=&  \langle Z_{\rm eff} \rangle^J + \sum_{L=J+1}^{\infty}
\frac{B_Z}{(L+{\scriptstyle \frac{1}{2}})^2 } \nonumber \\  
& \approx & \langle Z_{\rm eff} \rangle^J + \frac{B_Z}{J+{\scriptstyle \frac{1}{2}} } \ , 
\label{Glebfit2}
\end{eqnarray}
and so determined $Z_{\rm eff}$ and $\delta$.  They
used the approximate identities in eqs.~(\ref{Glebfit1}) and (\ref{Glebfit2}) 
rather than explicitly evaluating the infinite sum.  The identities appear to 
have been derived as an approximation to the 
$\int^{\infty}_{J+1} (L+{\scriptstyle \frac{1}{2}})^{-2} \ dL$ integral.  
However, in equating the sum to the integral they implicitly assume a 
rectangle rule representation of the integral which is in error of 
5-10$\%$ for $J \in [7,10]$ (the net effect of this is that Gribakin and
Ludlow state that the increments decrease as 
$B/(L+{\scriptstyle \frac{1}{2}})^{n}$ 
but actually assume a $B/L^{n}$ decrease when evaluating the $J \to \infty$ 
correction).  A better 
approximation to the series is obtained by using a mid-point rule to
represent the integral.  Doing this leads to  
\begin{equation} 
\sum_{L=J+1}^{\infty} \frac{1}{(L+{\scriptstyle \frac{1}{2}})^n } \approx  \frac{1}{(n-1)(J+1)^{n-1} }  \ .   
\label{bettertail} 
\end{equation} 
This approximation is accurate to 0.1$\%$ for $n = 2$ and $J = 7$.  

It will be shown that a more serious problem with the Gribakin and Ludlow 
methodology is that eqs.~(\ref{Glebfit1}) and (\ref{Glebfit2}) cannot reveal 
whether the calculated $Z^{J}_{\rm eff}$ are deviating from the expected 
asymptotic form.  For example,  successive increments to either the phase 
shift or $Z_{\rm eff}$ have usually decreased more slowly with $J$ (for $
J$ ranging between 10 and 18) than indicated by eqs.~(\ref{pE})-(\ref{pZ}) 
\cite{bromley02a,bromley02b,bromley03a,novikov04a}.   Instead of having 
$p = 2$ or $p = 4$ the successive increments often gave slightly smaller values 
for $p$.  The approach adopted by Gribakin and Ludlow is insensitive to these 
deviations.
  
Saito has investigated the structure of the PsH and the
Ps-halogen systems with the CI method \cite{saito03a,saito05a}.  A 
Natural Orbital (NO) truncation algorithm based on the energy was used 
to reduce the dimensionality of the secular equations, thus making 
calculations on the heavier halogen atoms viable.  Besides using the 
inverse power series, Saito used the functional form    
\begin{equation}
\Delta X^{J} = 10^{-\alpha(\log_{10}J)^\beta+\gamma} \ ,  
\label{plog}
\end{equation}
to complete the partial wave sum for the annihilation rate.  This 
function was not based on any physical principles, and its usage was 
justified on the grounds that the increments were decreasing faster 
than $J^{-2}$.  However, it has been suggested that the annihilation 
rate increments were decreasing too quickly because the dimension of 
the radial basis used in the Ps-halogen calculations was simply too
small \cite{mitroy05i}.  So the rationale behind the usage of 
eq.~(\ref{plog}) is questionable.     
      
Some mention must be made of the difficulties associated with 
the slower convergence of the annihilation rate.  Consider  
the PsH system, a calculation with $J = 9$ gave 72$\%$ of the 
total annihilation rate \cite{bromley02a}.  If one doubled the size of 
$J$, then eq.~(\ref{pGamma}) suggests that the explicit calculation 
would only recover 86$\%$ of the total annihilation rate.  
And it would take a calculation with $J \approx 250$ to recover 
99$\%$ of the annihilation rate.  The situation is even more sobering 
when one considers that the annihilation rate converges faster for PsH 
(since it is the most compact) than for any other positron binding system. 

\section{Comparison of existing and new approaches to the partial wave extrapolation} 

\subsection{The different alternatives} 

In order to expose the strengths and deficiencies of existing approaches, 
very large calculations have been performed on three mixed electron-positron
systems.  These are the $e^+$-H scattering system for the $\ell = 0$ 
partial wave, and the bound PsH and $e^+$Cu systems.  It will be seen that 
the typical calculations on these real-world systems do not agree perfectly 
with the leading order asymptotic form given by Gribakin and Ludlow, i.e. 
eqs.~(\ref{pE}) - (\ref{pZ}).  Accordingly, six different extrapolation methods 
for determining the $J \to \infty$ correction were tested.  These were:

\textbf{Method $p$}.  The successive increments to all quantities are assumed to 
obey an
\begin{equation} 
X^J =  \frac{A_X}{(J+{\scriptstyle \frac{1}{2}})^{p} }  \ ,  
\label{Xgen} 
\end{equation} 
law with the exponent $p$ determined from eq.~(\ref{pdef}).    
The value of $p$ derived from three successive calculation of 
$X^{J-2}$,$X^{J-1}$ and $X^J$ is given by    
\begin{equation}
p =   \ln \left(  \frac {\Delta X^{J-1}}{\Delta X^J} \right) \biggl/ 
      \ln \left( \frac{J+{\scriptstyle \frac{1}{2}}}{J-{\scriptstyle \frac{1}{2}}} \right) \ .  
\label{pdef} 
\end{equation}
(Note, in previous works we have used a $J^{-p}$ series 
\cite{mitroy99c,bromley02a,bromley02b,bromley03a}).  
The notations $p_E$, $p_{\Gamma}$, $p_{\delta}$ and $p_Z$ are used
to denote the exponents derived from the partial wave expansions 
of the energy, annihilation rate, phase shift and $Z_{\rm eff}$ 
The discrete sum over $L$ in (\ref{XJ2}) is done explicitly up 
to $J = 200$.  The remainder of the sum is then estimated 
using eq.~(\ref{bettertail}).        

\textbf{Method $p_{\mathrm{av}}$}.  This is based on Method $p$.  Three 
successive calculations for $(J\!-\!2)$, $(J\!-\!1)$ and $J$ are once again 
used with eqs.~(\ref{Xgen}) and (\ref{pdef}) to determine an initial 
estimate of $p_0$.  Then, $p$ is set to the average of $p_0$ and the 
expected value of either 2 or 4.   This method makes an admittedly 
crude attempt to correct method $p$ in those cases where $p_{\delta}$ 
and $p_Z$ were significantly different from 4 and 2 \cite{bromley02b}.  
Once $p$ has been fixed, eq.~(\ref{Xgen}) can then be 
used to determine $A_X$ and the discrete sum over $L$ in (\ref{XJ2}) 
is done explicitly up to $J = 200$.  The remainder of the sum is then 
estimated using eq.~(\ref{bettertail}).        

\textbf{Method GL}.  The relations eq.~(\ref{Glebfit1}) and (\ref{Glebfit2}) 
are assumed to be exact.  The two largest values of $\langle X \rangle ^J$ 
are used to determine $\langle Z_{\rm eff} \rangle^{\infty}$ and 
$\langle \delta \rangle^{\infty}$.  This method mimics the procedure 
adopted by Gribakin and Ludlow \cite{gribakin04b}.  

\textbf{Method I}.  The functional form 
\begin{equation} 
\Delta X^J = \frac{B_X}{(J+{\scriptstyle \frac{1}{2}})^n} 
\label{Lseries1} 
\end{equation} 
is assumed to apply and the $\Delta X^{J}$ increment 
are used to determine $B_X$.  The exponent $n$ is set to $2$ for 
the annihilation rate and $4$ for the energy or phase shift.  
The discrete sum over $L$ in (\ref{XJ2}) is done explicitly up to 
$J = 200$ and beyond that point eq.~(\ref{bettertail}) is used.        
This method has similarities with the GL method.  

\textbf{Method II}.  The functional form 
\begin{equation} 
\Delta X^J = \frac{B_X}{(J+{\scriptstyle \frac{1}{2}})^n} 
   + \frac{C_X}{(J+{\scriptstyle \frac{1}{2}})^{n+1}} \ , 
\label{Lseries2} 
\end{equation} 
is assumed to apply and the $\Delta X^{J}$ and $\Delta X^{J-1}$ increments are 
used to determine $B_X$ and $C_X$.  The second term in eq.~(\ref{Lseries2}) comes 
from 3rd-order perturbation theory \cite{hill85a,kutzelnigg92a,ottschofski97a}.    
The exponent $n$ is set to $2$ for 
the annihilation rate and $4$ for the energy or phase shift.  
The discrete sum over $L$ in (\ref{XJ2}) is done explicitly up to 
$J = 200$ and beyond that point eq.~(\ref{bettertail}) is used.        

\textbf{Method III}.  The functional form 
\begin{equation} 
\Delta X^J = \frac{B_X}{(J+{\scriptstyle \frac{1}{2}})^n} 
   + \frac{C_X}{(J+{\scriptstyle \frac{1}{2}})^{n+1}}
   + \frac{D_X}{(J+{\scriptstyle \frac{1}{2}})^{n+2}} \ , 
\label{Lseries3} 
\end{equation} 
is assumed to apply and the $\Delta X^{J}$, $\Delta X^{J-1}$ and 
$\Delta X^{J-2}$ increments are 
used to determine $B_X$, $C_X$ and $D_X$.   Other particulars are the same as 
those for Methods I and II. 
 
\textbf{Method S}.  The functional form adopted by Saito, eq.~(\ref{plog}) is 
used.   The parameters $\alpha$, $\beta$ and $\gamma$ are determined
from $\Delta X^{J-2}$, $\Delta X^{J-1}$ and $\Delta X^J$.  Then 
the series is completed by summing to $J=2000$.  

\subsection{The $e^+$-H scattering system} 

The CI-Kohn method has already been used to generate phase shift 
and annihilation parameter data for the $e^+$-H scattering system 
\cite{bromley03a}.  New calculations with a radial basis 
set (i.e. the number of LTOs per $\ell$) of increased dimensionality 
have been done in order to minimize the influence of the radial 
basis upon any conclusions that are drawn.  The present investigation 
examines $s$-wave $e^+$H scattering at $k = 0.4 \ \ a_0^{-1}$.   

\vspace{0.2cm}
\begin{table}[th]
\caption[]{  \label{tabpH}
Results of CI-Kohn calculations for $s$-wave $e^+$-H scattering at 
$k = 0.4 \ \ a_0^{-1}$.   The variational and close-coupling  
data in the last three rows are taken from calculations which have 
basis functions which explicitly depend on the electron-positron 
distance.  The $J \rightarrow \infty$ limits were taken at $J = 12$ 
using the various extrapolation methods as described in the text.  
The value of $p$ were determined at $J = 12$ using eq.~(\ref{pdef}).  
The previous CI-Kohn estimates \cite{bromley03a} include the 
$J \to \infty$ correction as determined at that time.  
}
\vspace{0.2cm}
\begin{ruledtabular}
\begin{tabular}{lcc} 
$J$   &  $\langle \delta \rangle^J$ (radians) &  $\langle Z_{\rm eff} \rangle^J$ \\
 \hline 
0  &   -0.1992097664    &   0.4529537597  \\ 
1  &   -0.01002215705   &   1.018467362   \\ 
2  &    0.05339800596   &   1.466777767   \\ 
3  &    0.08175604707   &   1.798430751   \\ 
4  &    0.09626735469   &   2.043817341   \\ 
5  &    0.1043829191    &   2.228559914   \\ 
6  &    0.1092338304    &   2.370688616   \\ 
7  &    0.1122907225    &   2.482386967   \\ 
8  &    0.1143023638    &   2.571897852   \\ 
9  &    0.1156750002    &   2.644890677   \\ 
10 &    0.1166408472    &   2.705331457   \\ 
11 &    0.1173386269    &   2.756059864   \\ 
12 &    0.1178543969    &   2.799146079   \\ \hline  
 \multicolumn{3}{c}{$J \rightarrow \infty$ limits}  \\ 
 $p$       &  3.6248  & 1.9583  \\
Method $p$       & 0.1200652 & 3.34020 \\
Method $p_{av}$  & 0.1199020 & 3.32823  \\
Method GL        & 0.1196691 & 3.29464  \\
Method I         & 0.1197593 & 3.31675  \\
Method II        & 0.1199484 & 3.32750  \\
Method III       & 0.1199456 & 3.30092  \\
Method S         & 0.1198834 & 3.21989  \\
 \multicolumn{3}{c}{Other calculations}  \\ \hline  
CI-Kohn $J \rightarrow \infty$ \cite{bromley03a} & 0.1198 & 3.232   \\
Optical potential \cite{bhatia71,bhatia74b}& 0.1201 & 3.327  \\
Variational \footnotemark[1] \cite{vanreeth97a,vanreeth98a,gribakin04b} & 0.1198 & 3.407  \\
Close Coupling \cite{mitroy95a,ryzhikh00a} & 0.1191 & 3.332   \\
\end{tabular}
\end{ruledtabular}
\footnotetext[1] {The variational results of van Reeth {\em et.al.} 
\cite{vanreeth97a,vanreeth98a} are only given in tabular form in 
\cite{gribakin04b}.}    
\end{table}

The largest calculation for $e^+$-H included a minimum of 30 
LTOs per $\ell$ with additional LTOs included at small $\ell$.     
Special attention was given to the $\ell = 1$ positron basis 
since this channel is responsible for the long-range polarization 
potential.  The dimension of the LTO basis was 80 in this case.  
All radial integrations were taken to 729 $a_0$ on a composite 
Gaussian grid.  The earlier calculations of Bromley and Mitroy 
\cite{bromley03a} with a minimum of 17 LTOs per $\ell$ will be 
presented for comparison.  Table \ref{tabpH} gives the $e^+$-H 
phase shift and $Z_{\rm eff}$ for $s$-wave scattering at 
$k = 0.4$ $a_0^{-1}$ up to $J = 12$.  The larger basis will be 
referred to as basis 2 while the older basis will be named 
basis 1.    

\begin{figure}
\centering
\includegraphics[width=8.0cm,angle=0]{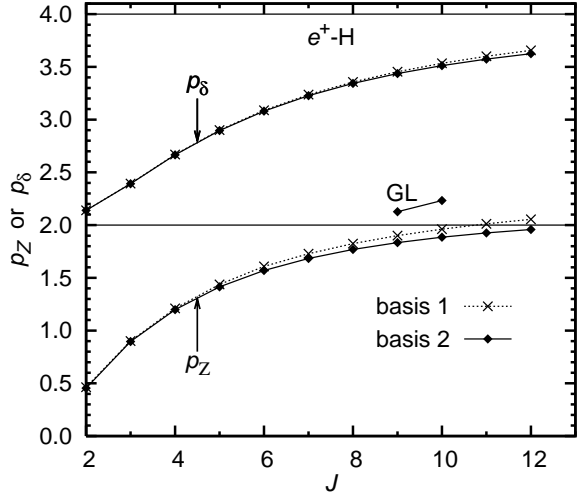}
\caption[]{
The exponents $p_{\delta}$ and $p_Z$ as a function of $J$ for 2 different 
CI-Kohn calculations of $s$-wave $e^+$-H scattering at $k = 0.4 \ a_0^{-1}$.
The short curve labeled GL was plotted using $\langle Z_{\rm eff} \rangle^J$  
data of GL \cite{gribakin04b}.  
}
\label{pHp}
\end{figure}

First of all, the values of $p_{\delta}$ and $p_Z$ 
derived from the earlier \cite{bromley03a} and present CI-Kohn 
calculations are shown together in Figure \ref{pHp} as a function of $J$.    
This figure tests whether eqs.~(\ref{pdelta}) and (\ref{pZ}) 
describe the behavior of a real-world calculation.  

Neither $p_{\delta}$ or $p_Z$ are within 1$\%$ of the expected 
asymptotic value at $J = 12$ and both are approaching the expected
asymptotic limit from below.  The two calculations give almost
exactly the same $p_{\delta}$ while the larger calculation tends
to give smaller values of $p_Z$ with the difference becoming greater  
as $J$ increases.  The increasing difference between $p_Z$ for the 
two calculations suggests that a converged calculation of 
$\Delta Z^J_{\rm eff}$ needs an increasingly larger radial basis 
as $J$ increases.  This point is addressed in more detail later.   
In most of the calculations we have performed, the values of 
$p$ derived from eq.~(\ref{pdef}) have been slightly smaller than 
the expected value at $J \approx 10$ 
\cite{bromley02a,bromley02b,bromley02d,novikov04a}.     
We also note that in all the calculations we have so far performed, 
the values of $p$ increase steadily (once the broad features 
of the physical system have been achieved) as $J$ increases.   
While the asymptotic increments to $\delta$ and $Z_{\rm eff}$ 
do not agree exactly with eqs.~(\ref{pdelta}) and (\ref{pZ}),  
the observed trends do appear to be consistent with 
their derived limits.     

\begin{figure}
\centering
\includegraphics[width=8.0cm,angle=0]{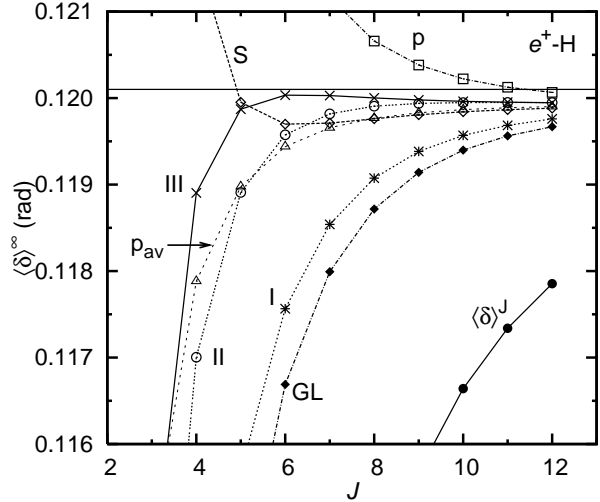}
\vspace{0.1cm}
\caption[]{
The extrapolated $J \rightarrow \infty$ limit of the phase shift 
(in radians) $\delta$ for $e^+$-H scattering at $k = 0.4 \ a_0^{-1}$ 
as a function of $J$.  The horizontal solid line shows the phase shift 
of Bhatia {\em et.al.} \cite{bhatia71}.   The phase shift without any 
$J \rightarrow \infty$ correction is given by the curve labeled 
$\langle \delta \rangle^J$.   
}
\label{Hphaseinf}
\end{figure}

\begin{figure}
\centering
\includegraphics[width=8.0cm,angle=0]{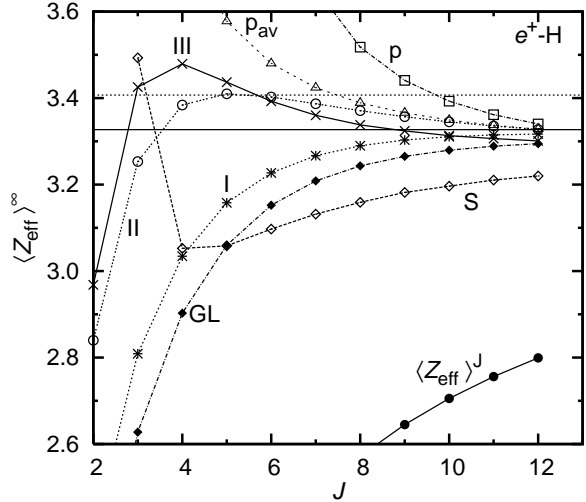}
\vspace{0.1cm}
\caption[]{
The extrapolated $J \rightarrow \infty$ limit using the different 
methods to complete the partial wave series for the $e^+$-H 
$Z^J_{\rm eff}$ at $k = 0.4 \ a_0^{-1}$ a function of $J$.   The 
horizontal solid line shows the $Z_{\rm eff}$ of Bhatia {\em et.al.} 
\cite{bhatia74b} while the horizontal dotted line shows the $Z_{\rm eff}$ 
of van Reeth {\em et.al.} \cite{vanreeth97a,vanreeth98a,gribakin04b}. 
}
\label{HZinf}
\end{figure}

Figures \ref{Hphaseinf} and \ref{HZinf} show the behavior of the 
extrapolated $\delta$ and $Z_{\rm eff}$ as a function of $J$ for 
some of the different extrapolations.  Table \ref{tabpH} 
gives estimates of $\delta$ and $Z_{\rm eff}$ using
the calculated values at the largest possible $J$ value (i.e. 12) 
to determine the $J \to \infty$ corrections.  

There is one problem in interpreting the results of Figures 
\ref{Hphaseinf} and \ref{HZinf} and Table \ref{tabpH}.  The
exact value of $Z_{\rm eff}$ is imprecise at the level of 2-3$\%$.  
The old calculation of Bhatia \textit{et.al.} using $r_{ij}$ co-ordinates 
gives 3.327 \cite{bhatia71,bhatia74b}, the close coupling calculation by 
Ryzhikh and Mitroy using the $T$-matrix method gives 3.332 
\cite{mitroy95a,ryzhikh00a}, while the variational calculation 
of van Reeth {\em et.al.} gives 3.407 
\cite{vanreeth97a,vanreeth98a,gribakin04b}.  
There is also some scatter in the estimates of the phase shifts, 
but the degree of difference between the Bhatia {\em et.al.} 
and van Reeth {\em et.al.} phase shifts is only 0.3$\%$.   
(the $T$-matrix phase shift is only expected to have an accuracy
of about 1$\%$ \cite{mitroy95a}).

Figure \ref{Hphaseinf} shows that inclusion of the $J \to \infty$ correction
leads to greatly improved estimates of $\langle \delta \rangle^{\infty}$.
In terms of their 
impact, the methods belong to 3 operational classes.  Firstly, Method $p$ 
consistently gives the largest values of $\langle \delta \rangle^{\infty}$.
Fixing $p_{\delta}$ to a set value at a finite $J$ inevitably 
results in the $J \to \infty$ correction being overestimated.
For example, using $p_{\delta} =  3.343$ at $J = 8$ to fix $p_{\delta}$ 
for all $J$ results in increments to $\Delta \delta^J$ that do not 
decrease quickly enough.  Methods I and GL, on the other hand, tend 
to underestimate the size of the $J \to \infty$ correction since 
$p_{\delta} = 4$ is fixed prior to the sequence of 
$\Delta \delta^J$ increments achieving the expected
$(J+{\scriptstyle \frac{1}{2}})^{-4}$ form.  

Those methods which attempt to allow for deviations from the
leading order behavior, namely Methods II, III and
$p_{av}$  approach the expected $J \to \infty$ limit much earlier.
Indeed, their estimates of $\langle \delta \rangle^{\infty}$ differ
by less than 0.5$\%$ at $J = 6$.  Table \ref{tabpH} reveals that
these 3 methods give $\langle \delta \rangle^{\infty}$ estimates 
that differ by less than 0.1$\%$ at $J = 12$.  Of the 3 alternatives, 
the 3-term asymptotic series, namely Method III, seems to possess the 
best convergence properties.   Method S also appears to give a reasonable  
estimate of $\langle \delta \rangle^{\infty}$ when $J \ge 6$.

The tabulated estimates of $\langle \delta \rangle^{\infty}$ at 
$J = 12$ reflect the discussions in the above paragraph.  Method
$p$ gives the largest phase shift while Methods GL and I give
the smallest phase shifts.  The maximum difference between any of 
the phase shifts is only 0.3$\%$ since the net effect of the 
$J \to \infty$ contribution to the phase shift is only 2$\%$.

Figure \ref{HZinf} for $Z_{\rm eff}$ shows some features in common 
with Figure \ref{Hphaseinf}. Once again, application of a $J \to \infty$ 
correction is seen to give much improved estimates of the 
$\langle Z_{\rm eff} \rangle^{\infty}$ limit.  Method $p$ also gives 
the largest estimate of $Z_{\rm eff}$.  Method S gives values of 
$\langle Z_{\rm eff} \rangle^{\infty}$ that are generally the smallest, 
and Table \ref{tabpH} 
reveals that it gives a value that is 0.1 smaller than any of the 
other approaches at $J = 12$.   It is not surprising that a method 
based on a fitting function with no physical justification performs 
so poorly, and its reasonable estimate of $J \to \infty$ correction 
to the phase shift can be regarded as a numerical coincidence.
No further discussion of Method S will be made although values 
are reported in the later tables for reasons of completeness.    

A more detailed analysis and discussion of Figure \ref{HZinf} cannot be made 
until the convergence properties of the underlying radial basis 
are exposed.

\subsection{Convergence properties of the radial basis}

In addition to converging very slowly with $J$, the annihilation rate 
also converges slowly with respect to the number of radial basis 
functions since the actual wave function 
has a cusp at the electron-positron coalescence point.   
A previous CI investigation on helium in an $\ell = 0$ model 
indicates that the electron-electron $\delta$-function converged as 
$O(N^{-5/2})$ where $N$ is number of Laguerre orbitals \cite{mitroy06a}.    
And it has also been demonstrated that the relative accuracy of the 
electron-electron $\delta$-function increment for a given size radial 
basis decreased as $L$ increased \cite{bromley06a}. 

\begin{figure}
\centering
\includegraphics[width=8.0cm,angle=0]{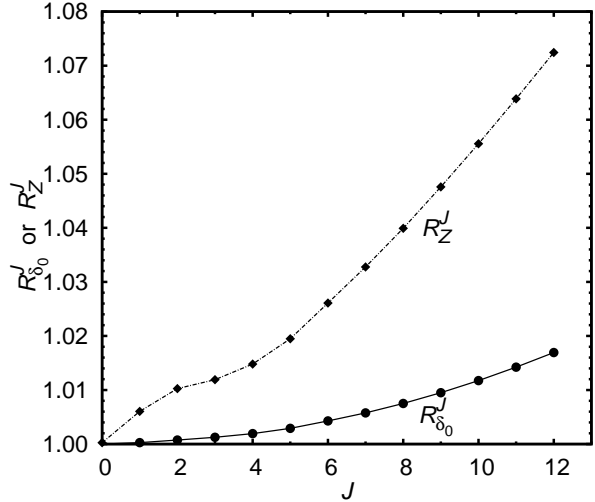}
\caption[]{
The ratio of the increments to $\langle \delta \rangle^J$  and
$\langle Z_{\rm eff} \rangle^J$ (refer to eq.(\ref{ratio})) for
$e^+$-H scattering at $k = 0.4$ $a_0^{-1}$ as a function of $J$ 
for the 17 and 30 LTO calculations.
}
\label{RdE}
\end{figure}

Some sample ratios can be used to illustrate these points.  The ratio 
$R^{J}_Z$ compares the two calculations of $\Delta Z^{J}_{\rm eff}$ 
for $e^+$-H scattering by determining the ratio for basis 1(17 LTOs) and 
basis 2(30 LTOs).  It is defined as 
\begin{equation}
R^{J}_Z = \frac{(\Delta Z^{J}_{\rm eff})_{30}} {(\Delta Z^{J}_{\rm eff})_{17}} \ . 
\label{ratio}
\end{equation}
A similar ratio, $R^{J}_{\delta}$ can be defined for the increment to the 
phase shifts.  A plot of these two ratios is given in Figure \ref{RdE},  
while  Table \ref{radialbasis} lists values of $R^{J}_{\delta}$ and $R^J_Z$ 
for some selected $J$ values.   
Figure \ref{RdE} clearly demonstrates the $\Delta Z_{\rm eff}^J$ 
converges more slowly that the phase shift increments at large $J$. 
First, the annihilation rate is more sensitive to the size of the 
radial basis than is the phase shift.  Second, the higher partial 
waves are more sensitive to the size of the radial basis than the 
lower partial waves.  For example there was a 4.0$\%$ increase in 
$\Delta Z^8_{\rm eff}$ between the basis 1 and basis 2 calculations 
while there was a larger 7.2$\%$ increase in $\Delta Z^{12}_{\rm eff}$.  
However, the increase in $\Delta \delta^{12}$ was only 1.7$\%$.   

Since the lack of completeness in the radial basis has the largest impact 
at high $J$, it will obviously affect the $J \to \infty$ correction.  For 
example, Method II gives $\Delta Z_{\rm eff}^{J > 12} = 0.4695$ for basis 
1.  For basis 2, the correction of $\Delta Z_{\rm eff}^{J > 12} = 0.5284$ 
is significantly larger.  

\begin{table}[th]
\caption[]{  \label{radialbasis}
Ratios of the partial wave increments to the energy (or phase shift), 
annihilation rate (or $Z_{\rm eff}$) taken from the two different 
basis sets used for each system.     
}
\vspace{0.2cm}
\begin{ruledtabular}
\begin{tabular}{lcc} 
$J$ &  $R_\delta$ &  $R_Z$ \\
 \hline 
\multicolumn{3}{c}{$e^+$-H }   \\
  6   &  1.0043  &  1.0261  \\
  8   &  1.0075  &  1.0399  \\
 10   &  1.0117  &  1.0556  \\   
 12   &  1.0169  &  1.0724  \\  \hline 
$J$   &  $R_E$   &  $R_{\Gamma}$ \\  \hline  
\multicolumn{3}{c}{$e^+$Cu }   \\
  8   &  1.0100  &  1.0288  \\
 12   &  1.0258  &  1.0552  \\
 16   &  1.0477  &  1.0869  \\  
\multicolumn{3}{c}{PsH }   \\
  4   &  1.0096  &  1.0478  \\
  6   &  1.0114  &  1.0825  \\
  9   &  1.0626  &  1.1357  \\
\end{tabular}
\end{ruledtabular}
\end{table}

The implications of these results can be seen by consideration of the 
Method II plot of $Z_{\rm eff}^{\infty}$ depicted in Figure \ref{HZinf}.  
This achieves a maximum value of 3.41 at $J = 5$, and then decreases 
until it is 3.328 at
$J = 12$.  The question to be addressed is whether the decrease from
$J > 6$ is due to convergence properties of the calculation with respect 
to $J$ or the convergence properties of the basis with respect to $N$,
the number of orbitals per $\ell$? 
Although there are seven estimates of $Z_{\rm eff}$ that lie between
3.30 and 3.34 in Table \ref{tabpH}, we believe that the true value of
$Z_{\rm eff}$ lies closer to 3.407 (the value of van Reeth {\em et.al.}) 
than to 3.327 (the value of Bhatia {\em et.al.}).  

This interpretation is supported by a crude estimate of the variational 
limit deduced from Figure \ref{RdE}. The assumption is made that the 
$\Delta Z_{\rm eff}^J$ increments converge as $O(N^{-5/2})$ 
\cite{mitroy06a}.  Consequently, one deduces that
the plotted $R^J_{Z}$ ratios comprise some 57$\%$ of the necessary
correction to the variational limit.  The variational limit for
any increment then estimated to be 
$\Delta Z_{\rm eff}^J \times (1 + 0.43 \times (R^J_{Z}-1)/0.57)$.
Applying this correction to the data in Table \ref{tabpH} gives  
$\langle Z_{\rm eff} \rangle^{\infty} = 3.410$ when Method II is used
to estimate the $J \to \infty$ limit (the actual correction of  
$\Delta Z_{\rm eff}^{J>12}$ of 0.578 was about 9$\%$ larger 
than the basis 2 value).
Determination of the variational limit for the phase shift
can also be done by assuming that the $\Delta \delta^J$
increments converge as $O(N^{-7/2})$ \cite{mitroy06a}.  In this case,  
the plotted $R^J_{\delta}$ ratios comprise some 76$\%$ of 
the correction to the variational limit and thus the final
estimates of the phase shift increments would be   
$\Delta \delta^J \times (1 + 0.24 \times (R^J_{\delta}-1)/0.76)$.
The Method II phase shift only increased by $1.3 \times 10^{-5}$ 
radian giving $\langle \delta \rangle^{\infty} = 0.12008$.

One point from Table \ref{tabpH} warrants special attention.  The
3-term asymptotic series, namely Method 3, gives a smaller   
$\langle Z_{\rm eff} \rangle^{\infty}$ than Method I!  This seems 
ridiculous given that $p = 1.9583$ at $J = 12$.  The tendency for 
the $\Delta Z_{\rm eff}^J$ increments to be systematically 
underestimated results in the corruption of the $B_Z$, $C_Z$ and
$D_Z$ coefficients extracted from the 3-term fit and renders
Method III unreliable for determination of 
$\langle Z_{\rm eff} \rangle^{\infty}$.

Even though data for only one system has been presented, it is possible
to make some general comments about the performance of the difference
methods since analysis of the $e^+$Cu and PsH data will confirm these 
conclusions (they are also compatible with the
results of large basis CI calculations of He \cite{bromley06a}).        

Since the $p_{\delta}$ and $p_Z$ exponents tend to be smaller than 4 and 2 
respectively at finite $J$, Method $p$ has an inherent tendency to overestimate 
the $J \to \infty$ corrections.  Application of this approach in the past 
has not resulted in any gross errors since the problems associated with 
fixing $p$ at values less than 2 and 4 tend to cancel out the errors  
associated with a radial basis of finite size $p$ \cite{bromley02d}.  
This method should only be applied in situations when the asymptotic
form of the expectation value under investigation is unknown.  

Method I generally underestimates the $J \to \infty$ correction.  
It gives a useful estimate of the $J \to \infty$ correction and should 
mainly be applied to give rough estimates for low precision
calculations.  Method GL can be regarded as a variety of Method I 
that happens to give inferior $J \to \infty$ corrections.    

Methods II and $p_{av}$ were seen to give $J \to \infty$ corrections
that were close to each other once the calculation reached a certain 
value of $J$.  Method II should be be preferred since 
it is founded in correct asymptotics.   

Method III seems to give the earliest reliable estimate of the phase
shift.  However, it should not be applied to the annihilation rate 
unless the radial basis is substantially larger than the present 
basis.  Method III should only be applied in situations where 
the underlying partial wave increments have an accuracy of better
than 1$\%$ and in addition the increments should vary smoothly and
not exhibit fluctuations.    
 
\subsection{The $e^+$Cu ground state} 

\begin{figure}
\centering
\includegraphics[width=8.0cm,angle=0]{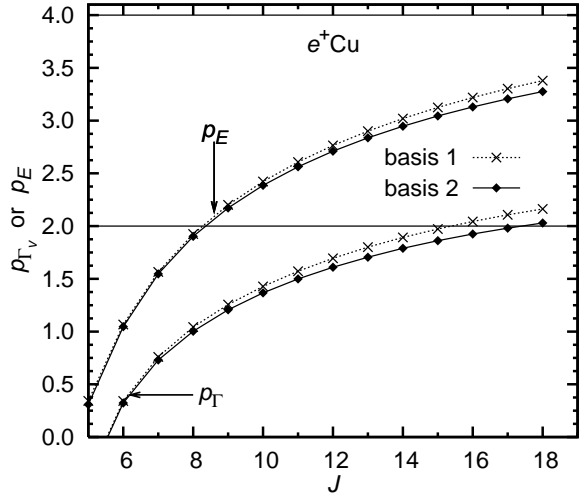}
\vspace{0.1cm}
\caption[]{
The exponents ($p_E$ and $p_{\Gamma}$) for 2 different CI calculations 
of $e^+$Cu as a function of $J$.   
}
\label{pCup}
\end{figure}

Table \ref{tabpCu} gives the $e^+$Cu binding energy and annihilation 
rate as a function of $J$ up to $J = 18$ for the calculation with
25 LTOs.  The table also includes values from a calculation with the
fixed core stochastic variation method (FCSVM) \cite{ryzhikh98f,bromley02e}.  
The FCSVM basis includes the electron-positron coordinate explicitly 
and is very close to convergence.  The FCSVM calculation uses a slightly
different model potential so it is not expected that the CI energy and
$\Gamma$ should be exactly the same. 
Figure \ref{pCup} displays $p_E$ and $p_{\Gamma}$ versus $J$ for two 
different CI calculations of the $e^+$Cu ground state.  One plot is 
derived from the earlier calculation of Bromley and Mitroy 
\cite{bromley03a} which included a minimum of 15 LTOs per $\ell$-value
(basis 1).  The present
calculation (basis 2) is much larger with a minimum of 25 LTOs per $\ell$ 
value (note, more than 25 LTOs were included for $\ell$ = 0, 1 and 2 since 
these make the largest contribution to the energy and annihilation rate).   

\begin{table}[th]
\caption[]{ Results of CI calculations on $e^+$Cu versus $J$ for the 
energy and annihilation rate for a series of $J$.  The binding energy 
of the positron to neutral Cu is denoted $\varepsilon$ (the energy
of Cu with respect to the core was -0.28394227 Hartree \cite{bromley02e}).  
The spin-averaged $2\gamma$ annihilation rates are given for the core 
($\Gamma_c$) and valence ($\Gamma_v$) electrons.
The results in the row $\infty$ include the $J \rightarrow \infty$ 
correction evaluated at $J = 18$ using the various methods 
described in the text.
}
\label{tabpCu}
\vspace{0.5cm}
\begin{ruledtabular}
\begin{tabular}{lccc}
$J$ & $\varepsilon$ & $\Gamma_c$ & $\Gamma_v$ \\   
   & Hartree &  $10^9$ sec$^{-1}$  &  $10^9$ sec$^{-1}$ \\ \hline  
0 &  -0.00112467 &   0.000289 & 0.000132  \\
 1 &  -0.00080292 &   0.001443 & 0.001692  \\
 2 &  -0.00037356 &   0.004818 & 0.009728  \\
 3 &  0.00031179  &   0.011213 & 0.033656  \\
 4 &  0.00111995  &   0.017605 & 0.070360  \\
 5 &  0.00187958  &   0.022223 & 0.110121  \\
 6 &  0.00251736  &   0.025271 & 0.147793  \\
 7 &  0.00302852  &   0.027272 & 0.181727  \\
 8 &  0.00343136  &   0.028611 & 0.211657  \\
 9 &  0.00374774  &   0.029526 & 0.237830  \\
10 &  0.00399692  &   0.030168 & 0.260656  \\
11 &  0.00419429  &   0.030627 & 0.280571  \\
12 &  0.00435174  &   0.030963 & 0.297984  \\
13 &  0.00447832  &   0.031212 & 0.313256  \\
14 &  0.00458086  &   0.031402 & 0.326694  \\
15 &  0.00466456  &   0.031547 & 0.338564  \\
16 &  0.00473338  &   0.031660 & 0.349087  \\
17 &  0.00479037  &   0.031749 & 0.358454  \\
18 &  0.00483788  &   0.031821 & 0.366821  \\ \hline
 \multicolumn{4}{c}{$J \rightarrow \infty$}  \\ 
$p$ &  3.2751      & 4.0511   & 2.0295    \\
Method $p$                & 0.0052012  & 0.032219 & 0.51307 \\
Method $p_{\mathrm{av}}$ & 0.0051481 &          & 0.51526  \\
Method GL                 & 0.0050998  &          & 0.51325  \\
Method I                  & 0.0051080 &          & 0.51751  \\
Method II                 & 0.0051584 &          & 0.51529  \\
Method III                & 0.0051603 &          & 0.49364 \\
Method S                  & 0.0051394  &          & 0.46398  \\ \hline 
              \multicolumn{4}{c}{Earlier calculations}   \\
FCSVM    \cite{ryzhikh98f,bromley02e}  & 0.005597    & 0.0339  &    0.544  \\
CI, $J = 18$ \cite{bromley02e}             & 0.004786    &  0.03173 & 0.35499 \\
CI, $J \to \infty$ \cite{bromley02e}       & 0.005117   & 0.0321 & 0.4744  \\
\end{tabular}
\end{ruledtabular}
\end{table}

\begin{figure}
\centering
\includegraphics[width=8.0cm,angle=0]{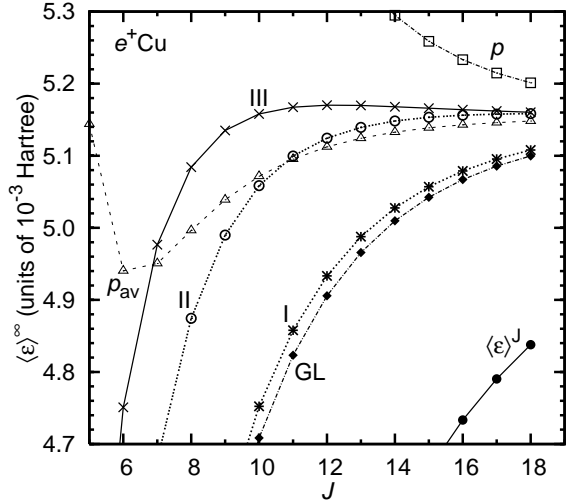}
\vspace{0.1cm}
\caption[]{
The $e^+$Cu binding energy as a function of $J$.  The different curves
use different algorithms to estimate the $J \rightarrow \infty$ 
correction as discussed in the text. 
}
\label{CupEinf}
\end{figure}

\begin{figure}
\centering
\includegraphics[width=8.0cm,angle=0]{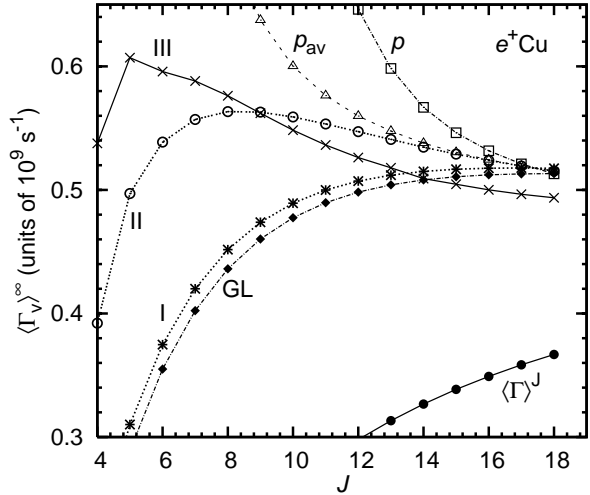}
\vspace{0.1cm}
\caption[]{
The $e^+$Cu annihilation rate (in units of $10^9$ s$^{-1}$) as a 
function of $J$.  The different curves use different algorithms 
to estimate the $J \rightarrow \infty$ correction as discussed 
in the text. 
}
\label{CupGinf}
\end{figure}

The plots of $p_E$ and $p_{\Gamma}$ against $J$ for $e^+$Cu are similar 
to the plots of $p_{\delta}$ and $p_Z$ for $e^-$-H scattering.  Both 
exponents are generally smaller than the expected asymptotic limits 
but steadily increase as $J$ increases.  The actual value of 
$p_{\Gamma}$ at $J = 18$, namely 2.030, was marginally larger than 
the expected asymptotic limit of $p_{\Gamma} = 2$.  The estimates of  
$\langle \varepsilon \rangle^{\infty}$ and $\langle \Gamma \rangle^{\infty}$ 
as a function of $J$ are shown in Figures \ref{CupEinf} and \ref{CupGinf}.  

Table \ref{tabpCu} reveals an 8$\%$ increase in $\Gamma$ when compared 
with the earlier CI calculation value of $\Gamma$, namely 
0.474 $\times 10^9$ s$^{-1}$ \cite{bromley02e}.  This is a consequence 
of the bigger radial basis used in the present work.   The selected 
values of $R^J_{\Gamma}$ listed in table \ref{radialbasis} reveal an 
8.7$\%$ increase in $\Delta \Gamma^{16}$ for basis 2 compared to
basis 1.  It is expected that further increases in the radial basis 
would eventually lead to a $p_{\Gamma}$ that was smaller than 2.0 
at all $J = 18$.  Again it is noticed that $R^J_{\Gamma}$ (and
$R^J_E$) increase with increasing $J$.  

The estimates of the annihilation rate in table \ref{tabpCu} 
are all very close together (with the exception of Method III). 
This occurs because $p_{\Gamma}= 2.030$ is very close
to the expected value of 2.0.  The variations between the different
approaches are largely concerned with taking care of the deviations 
from the $p_{\Gamma} = 2$ behavior, and with a minimal deviation 
at $J =18$, one should expect minimal differences between the final
results.  Method III is the least accurate (discounting Method S) since
it is the most susceptible to the inaccuracies in $\Delta \Gamma^{J}$ 

Figure \ref{CupGinf} shows that Method $p$ systematically overestimates  
$\langle \Gamma \rangle^{\infty}$ at smaller values of $J$.  Method I,  
on the other hand generally gives the smallest estimates of the 
$\langle \Gamma \rangle^{\infty}$ and is consistently too small at lower $J$.  

Method III obtains a reasonable estimate of $\langle \varepsilon \rangle^{\infty}$ 
the quickest.  Beyond $J > 12$ the Method III binding energy does decrease slightly.  
This may be due to slower convergence of the radial basis at higher
$J$.   Methods II and $p_{\mathrm{av}}$ give roughly equal estimates
of $\langle \varepsilon \rangle^{\infty}$ for $J > 10$, and the spread 
between the Methods II, III and $p_{\rm av}$ estimates of 
$\langle \varepsilon \rangle^{\infty}$ is only 0.2$\%$ at $J = 18$.

Method $p$ gives the largest estimate of $\langle \varepsilon \rangle^{\infty}$ 
for all $J$ shown in Figure \ref{CupEinf}, while Methods I and GL give the
smallest values (with Method GL once again being worse that Method I). 
It is worth mentioning that Methods II, III and $p_{\rm av}$ are all
roughly constant after $J = 16$ while Methods I, GL and $p$ are
still increasing or decreasing.

In summary, the totality of information in Table \ref{tabpCu} and Figures  
\ref{pCup}, \ref{CupEinf} and \ref{CupGinf} is very reminiscent 
of the situation for $e^+$-H scattering and is consistent with the
conclusions derived from the analysis of $e^+$-H scattering. 

\subsection{The PsH ground state} 

\begin{table}[th]
\caption[]{ Results of CI calculations on PsH for orbital bases with 
$L_{\rm int} = 4$ and for a series of $J$.  The total number of electron 
and positron orbitals are denoted by $N_{\rm orb}$, the total number
of configurations is given by $N_{\rm CI}$, and the LTO exponent for
$\ell = J$ is listed in the $\lambda$ column.  The 3-body energy of 
the PsH system in Hartree is denoted by $E(\text{PsH})$ and the 
$\Gamma$ is given in $10^9$ sec$^{-1}$.  
The $J \rightarrow \infty$ extrapolations were carried out at $J = 13$.
}
\label{PsH}
\vspace{0.5cm}
\begin{ruledtabular}
\begin{tabular}{lcccccc}
$J$ & $\lambda$ & $N_{\rm orb}$ & $N_{CI}$ & $E(\text{PsH})$ & $\Gamma$  \\ \hline 
 0 & 2.10 &  17 &   2601 & -0.69133618 &  0.374196  \\
 1 & 2.26 &  33 &   9265 & -0.74705969 &  0.782256  \\
 2 & 2.36 &  48 &  22810 & -0.76620031 &  1.080456  \\
 3 & 2.46 &  63 &  44650 & -0.77514128 &  1.292538  \\
 4 & 2.52 &  78 &  78640 & -0.77995286 &  1.448216  \\
 5 & 2.72 &  93 & 120265 & -0.78274494 &  1.566206  \\
 6 & 2.93 & 108 & 165265 & -0.78449597 &  1.658344  \\
 7 & 3.13 & 123 & 213415 & -0.78565220 &  1.732023  \\
 8 & 3.34 & 138 & 263365 & -0.78644538 &  1.792061  \\
 9 & 3.56 & 153 & 314890 & -0.78700639 &  1.841756  \\
10 & 3.75 & 168 & 366415 & -0.78741330 &  1.883366  \\
11 & 3.95 & 183 & 417940 & -0.78771485 &  1.918595  \\
12 & 4.15 & 198 & 469465 & -0.78794247 &  1.948689  \\
13 & 4.35 & 213 & 520990 & -0.78811707 &  1.974632   \\ \hline 
                           \multicolumn{6}{c}{$J \to \infty$ limits}   \\
\multicolumn{4}{l}{$p$}  & 3.44547     & 1.92375    \\ 
\multicolumn{4}{l}{Method $p$}               & -0.7889974  & 2.3352 \\
\multicolumn{4}{l}{Method $p_{\mathrm{av}}$}& -0.7888995  &  2.3232  \\
\multicolumn{4}{l}{Method GL}                & -0.7887894  & 2.2988  \\
\multicolumn{4}{l}{Method I}                 & -0.7888198 & 2.3121  \\
\multicolumn{4}{l}{Method II}                & -0.7889218 & 2.3225 \\
\multicolumn{4}{l}{Method III}               & -0.7889231 & 2.2915  \\
\multicolumn{4}{l}{Method S}                 & -0.7888843 & 2.2203  \\
                           \multicolumn{6}{c}{Other calculations and earlier CI calculations}   \\
\multicolumn{4}{l}{SVM \cite{mitroy06d}}         & -0.78919674 &  2.4712  \\  
\multicolumn{2}{l}{$J\! = \! 9$, $L_{\rm int} \! = \! 4$ \cite{bromley02a}} & 90/91  &  63492  & -0.7866818 &  1.7903  \\
\multicolumn{2}{l}{$J \! = \! 9$, $L_{\rm int} \! = \! 9$ \cite{bromley02a}} & 90/91    &  95324  & -0.7867761  & 1.7913  \\
\multicolumn{2}{l}{$J = 9$, \cite{saito03a}} & $\infty$    & $\infty$   & -0.786949 & 1.8230  \\
\end{tabular}
\end{ruledtabular}
\end{table}

The best two electron system for validation purposes is the positronium-hydride 
(PsH) system since its properties are very well known as a result of previous 
investigations \cite{ryzhikh99a,yan99a,usukura98,mitroy06d}.  The stochastic 
variational method (SVM) expectation values 
listed in Table \ref{PsH} are taken from a new calculation with 1800 ECGs. 
The energy of this wave function, $E = -0.7891674$ Hartree  
is the best PsH energy so far reported \cite{mitroy06d}.  

The orbital basis used in the present calculation was about twice as large 
as that used in previous calculations \cite{bromley02a,saito03a}.  
The number of radial functions per $\ell$ was 15 with the exception of 
$\ell = 0$ and $\ell = 1$ where 17 and 16 functions respectively were used.  
The largest $J$ was 13 while $L_{\rm int}$ was set to 4.  The basis functions 
for each $\ell$ used a common exponent that had been energy optimized during
some preliminary and smaller calculations.  It must be emphasized that 
choosing a common $\lambda$ for both electron and positron states was not 
arbitrary but was a consequence of energy optimization process.  The results 
of the CI calculation are listed in Table \ref{PsH}.     

\begin{figure}
\centering
\includegraphics[width=8.5cm,angle=0]{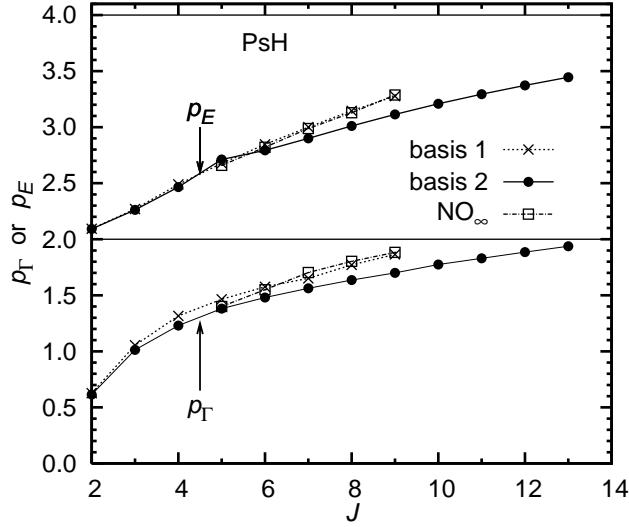}
\vspace{0.1cm}
\caption[]{
The exponents $p_{E}$ and $p_{\Gamma}$ as a function of $J$ for 2 different 
CI calculations of PsH. Both the basis 1 and basis 2 with $L_{\rm int} = 4$.  
The third data set was taken from the ``full CI'' calculations of Saito 
\cite{saito03a}.   
}
\label{pPsH}
\end{figure}

The variation of $p_E$ and $p_{\Gamma}$ with $J$ in Figure \ref{pPsH} reflects 
the behavior seen in Figures \ref{pHp} and \ref{pCup}. The values of $p$ are 
smaller than the predicted asymptotic limits and seem to be approaching the 
correct value.  Computational constraints mean that the dimensions of the 
radial basis,  e.g. 15 $e^+$ and $e^-$ LTOs per $\ell$, are smaller than 
those in the $e^+$-H and $e^+$Cu calculations.  The PsH radial basis is
further from convergence than the basis sets used for the equivalent 
calculations upon $e^+$-H and $e^+$Cu.   

Figure \ref{pPsH} gives $p$ values taken from the CI calculations of 
Saito \cite{saito03a}.  Saito used Natural Orbital techniques to reduce 
the dimension of the final diagonalization while using an orbital basis 
with $J = 9$ (this is about the same size as in \cite{bromley02a}). Saito 
estimated the variational limit at each $J$ and 
the Figure \ref{pPsH} curves were derived from this ``full CI limit'' 
calculation.  Although the Saito curves have irregularities, they 
exhibit a $p$ versus $J$ variation similar to the early Bromley 
and Mitroy calculation \cite{bromley02a} 

\begin{figure}
\centering
\includegraphics[width=8.5cm,angle=0]{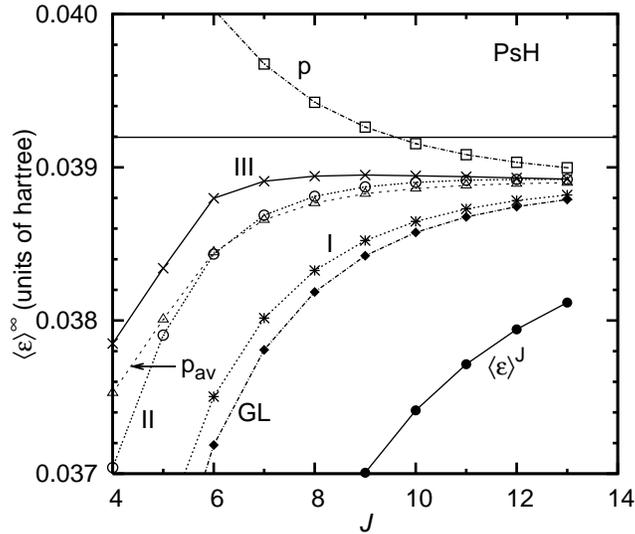}
\vspace{0.1cm}
\caption[]{
The PsH binding energy (in units of Hartree) with respect to the 
Ps+H threshold of -0.750 Hartree as a function of $J$.  The 
different curves use different algorithms to estimate the 
$J \rightarrow \infty$ correction as discussed in the text.  
The close to converged SVM energy \cite{mitroy06d} is shown as 
the horizontal line for comparison purposes.  
}
\label{PsHEinf}
\end{figure}

Figure \ref{PsHEinf} shows the variation of 
$\langle \varepsilon \rangle^{\infty}$ vs $J$.  Once again, the
3-term series, Method III achieves its asymptotic 
value at the smallest value of $J$.  Methods II and $p_{\rm av}$   
achieve their limiting values near $J = 10$.   Methods I and GL 
again tend to underestimate $\langle \varepsilon \rangle^{\infty}$ 
while Method $p$ overestimates $\langle \varepsilon \rangle^{\infty}$.
 
The best CI estimate of the PsH energy is obtained by adding an 
$L_{\rm int}$ correction of $9.5 \times 10^{-5}$ Hartree (the 
difference between the $L_{\rm int} = 4$ and $L_{\rm int} = 9$ 
energies \cite{bromley02a}) to the Method III $\varepsilon$ of $0.038923$ 
Hartree.  The resulting binding energy of $\varepsilon = 0.039018$ Hartree 
is only 0.38 $\%$ smaller than the SVM $\varepsilon$ of 0.0391674 Hartree.

\begin{figure}
\centering
\includegraphics[width=8.5cm,angle=0]{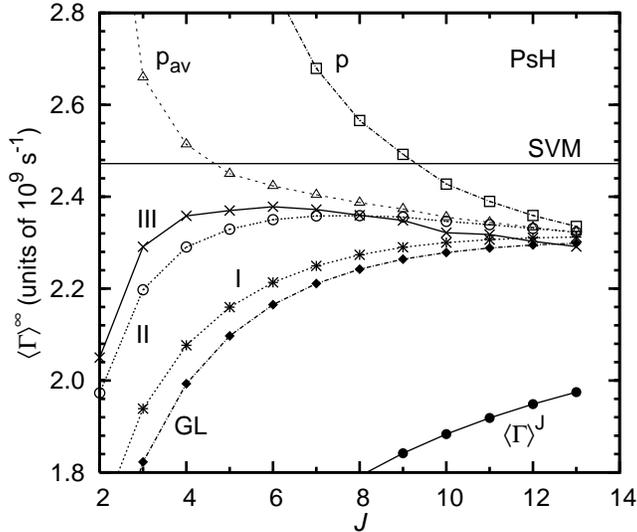}
\vspace{0.1cm}
\caption[]{
The PsH annihilation rate ($\langle \Gamma \rangle^{\infty}$ in units of 
10$^{9}$ s$^{-1}$) as a function of $J$.   The close to converged SVM 
annihilation rate is shown as the horizontal line.  
}
\label{PsHGinf}
\end{figure}

Figure \ref{PsHGinf} shows the annihilation rate versus $J$.  The
CI calculation does not converge to the SVM annihilation rate since  
a radial basis of 15 LTOs per $\ell$ is simply too small.  The 
$R^J_{\Gamma}$ entries in Table \ref{radialbasis} reveal a 13.5$\%$ 
increase in $\Delta {\Gamma}^{9}$ between the basis 1 and 
basis 2 calculations.  The conclusions that can be drawn from
Figure \ref{PsH} under such circumstances are somewhat limited.
But Method III is again susceptible to the accuracy of the
$\Delta \Gamma^J $ increments and again gives a final 
$\langle \Gamma \rangle^{\infty}$ that is smaller than 
Method I.  Methods II and $p_{\rm av}$ again give final estimates 
of $\langle \Gamma \rangle^{\infty}$ that are close together.  Method $p$ 
gives the largest estimate of $\langle \Gamma \rangle^{\infty}$ 
while Methods I and GL give the smallest.

Another problem with Method III arose from the usage of the Davidson
method to perform the matrix diagonalization.  This only gives values of 
$\langle \Gamma \rangle^J$ that are stable to 6-7 significant digits and
this leads to the irregularities in the $\langle \Gamma\rangle^{\infty}$
evident in Figure \ref{PsHGinf}.  The problem of decreased precision 
when using the Davidson algorithm had been previously noted in
CI calculations of helium \cite{bromley06a} and may be generic to
iterative matrix solvers.

\section{Comment on the scaling of the annihilation rate}  

There is one class of system that has not been studied in the 
present work, namely the close to threshold scattering of positrons 
from atoms that can bind a positron.  The behavior of the
$\Delta Z_{\rm eff}^J$ increments is complicated by a parametric 
dependence on the scattering length, $A$ 
\cite{mitroy02a,bromley02e,bromley03a}.  Once $J$ is
large enough to formally bind a positron, the magnitude of the
scattering length decreases as $J$ increases.   Since 
$Z_{\rm eff} \propto A^2$, the decrease in $A$ as $J$ increases 
impinges on the increase in $Z_{\rm eff}^J$ that would otherwise 
occur.   Indeed, one of the reasons why calculations on $e^+$Cu were 
originally taken to $J = 18$ was to minimize the disruption that the 
scattering length had on $Z_{\rm eff}$ \cite{bromley02e,bromley03a}.    

It was not worthwhile to try and analyze the behavior of the 
partial wave expansion for the $e^+$ + Cu scattering system with
its two major complications, the effect of the radial basis
set and, secondly, the effect of the $A$ versus $J$ variation.  
Such an investigation is best delayed until substantially larger
basis sets can be deployed. 

\section{Commentary on the work of Gribakin and Ludlow}

The problems caused by the slow convergence of $Z_{\rm eff}$ can be 
exposed by a detailed analysis of the recent Gribakin and Ludlow 
\cite{gribakin04b} 
calculation of the annihilation rate for $e^+$-H scattering.  This calculation 
used a variant of many-body perturbation theory (MBPT) to generate the
initial set of $\langle Z_{\rm eff}\rangle^J$ data.  They then did a fit to 
eq.~(\ref{Glebfit2}) over the $J = 7 \to 10$ interval and then plotted the
results of that fit under the assumption that this demonstrated that their 
data obeyed eq.~(\ref{Glebfit2}).  However, this procedure as applied by GL 
could hardly have been better suited to concealing deviations of the sequence 
of $\langle Z_{\rm eff}\rangle^J$ values from the leading order 
$(J+{\scriptstyle \frac{1}{2}})^{-2}$ term (note, in this section the
GL acronym refers to the results presented in \cite{gribakin04b}).  In effect, 
Figures 7 and 8 of GL are actually a demonstration of Taylor's theorem, 
namely that any continuous function will approximate a straight line if 
examined over a sufficiently small domain.   

\begin{figure}
\centering
\includegraphics[width=7.6cm,angle=0]{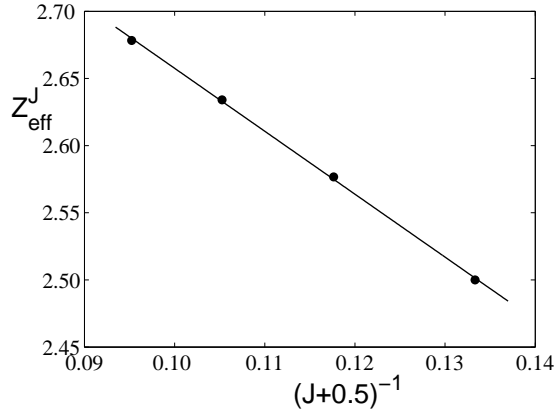}
\caption[]{
Plot of $Z^J_{\rm eff}$ vs $(J+{\scriptstyle \frac{1}{2}})^{-1}$ for
the synthetic data as discussed in the text.  The line shows the fit
to eq.~(\ref{Glebfit2}) while the $\bullet$ give the synthetic data points with the 
$\Delta Z^J_{\rm eff} \sim (J+{\scriptstyle \frac{1}{2}})^{-2.6}$  
dependence.  
}
\label{synthetic}
\end{figure}

To illustrate this point, consider a sequence of synthetic 
$\langle Z_{\rm eff} \rangle^J$ data generated by the  
prescription, $\langle Z_{\rm eff}\rangle^7 = 2.5$ and 
$\Delta Z^J_{\rm eff} = 20 \times (J+{\scriptstyle \frac{1}{2}})^{-2.6}$
(such a data sequence could be generated exactly from a 3-term expansion 
of eq.~(\ref{Zseries})).  A fit to this sequence was made using 
eq.~(\ref{Glebfit2}) and the results of that fit are shown in Figure 
\ref{synthetic}.   Even though the data were generated according to $p = 2.6$, 
a visual inspection (and it should be noted that Figure \ref{synthetic} is
much higher resolution than Figures 7 and 8 of GL) suggests that the data 
are in good agreement with a $p = 2$ power law decay whereas this
is certainly not the case.   

The $p_Z$ exponent derived from the recent GL calculation \cite{gribakin04b} 
is depicted in Figure \ref{pHp}.   Although GL assume $p_Z = 2$ when making 
the extrapolation, the actual exponent extracted from their sequence 
of $Z_{\rm eff}$ values is $p = 2.233$ at $J = 10$.  The present basis 2 
calculation gives $p_Z = 1.885$ at $J = 10$.  It is obvious that
the GL calculation overestimates the rate at which $\Delta Z^J_{\rm eff}$ 
increments are decreasing.  This implies that the GL estimates of 
$\Delta Z^J_{\rm eff}$ will be too small at higher $J$ and this is the
case.  GL get $\Delta Z^{10}_{\rm eff} = 0.0501$ while the CI-Kohn calculation 
gives $\Delta Z^{10}_{\rm eff} = 0.0604$.  Consequently, the GL calculation 
gives $\Delta Z^{J>10}_{\rm eff} = 0.48$, while the basis 2 CI-Kohn 
calculation gives $\Delta Z^{J>10}_{\rm eff} \approx 0.62$.  The assertion 
by GL that their calculation has converged to the region in which the 
$\Delta Z^J_{\rm eff} = B (J+{\scriptstyle \frac{1}{2}})^{-2}$ formula 
is valid is incorrect.  

Besides directly leading to less reliable extrapolation corrections, a related  
problem with the GL procedure is that it does not have the sensitivity to 
flag potential problems with the radial basis.  A plot of $p_Z$ versus $J$ 
that crosses the $p_Z = 2$ line is a good indicator of some inadequacy in the 
basis.   There is no indication that GL were aware that their 
$\Delta Z^{J}_{\rm eff}$ were decreasing much too quickly as $J$ increased, 
their statement that {\em The use of a $B$-spline basis means that fast 
convergence is achieved with respect to the number of states with a particular 
angular momentum} is difficult to reconcile with the present analysis.  However, 
it should be noted that do indicate that they could improve the quality of 
their answers by {\em ``pushing harder'' the numerics}.  

The tendency for the GL calculation to overestimate the convergence of
the annihilation rate increments probably does not arise from MBPT per
se, rather it most likely comes from the underlying single electron basis.
Besides the inherently slower convergence at higher $J$ mentioned
earlier, another possibility is due to the confinement of the basis to 
a box of radius 15 $a_0$.  Confining the basis in this way will result in mean
excitation energies (e.g. for predicting the multi-pole polarizabilities)
that will eventually increase as $\sim \ell^2$, where $\ell$ is the orbital
angular momentum, while for a real H atom the mean excitation energy for
any $\ell$ is less than 1.0 Hartree \cite{dalgarno56a}.  Thus the occupancy
of the higher $J$ orbitals, which contribute significantly to $Z_{\rm eff}$
will be inhibited, and successive $\Delta Z^{J}_{\rm eff}$ will decrease
too rapidly with increasing $J$.  

The relevance of these issues is best illustrated by a comparison with  
the exact value of $Z_{\rm eff}$ which will be taken to be 3.407 at 
$k = 0.4 \ a_0^{-1}$.  This is 0.295 larger than the GL value of 3.112.
The underestimation of the higher partial wave contribution in the GL 
calculation, estimated at $0.14=0.62-0.48$ is responsible for about 
50$\%$ of the existing discrepancy.    

The discrepancies of GL with the best calculations are are not that 
severe for $s$-wave scattering since imposition of the $p = 2$ 
condition for $J \ge 10$ prevents the inherent deficiencies in their 
$B$-spline basis from becoming too excessive.  Also $s$-wave $e^+$-H 
scattering system is certainly one of the easier positron annihilation 
calculations.  However, the inadequacy of the GL methodology manifests 
itself more severely in other positron annihilation situations.      
    
The CI expansion converges quicker for electron-positron annihilations 
that take place at small distances from the nucleus than for annihilations 
that take place at large distances \cite{mitroy02b,bromley02e,novikov04a}.  
The presence of the centrifugal barrier for $L > 0$ scattering leads to 
the electron-positron annihilations occurring further from the nucleus.  
Consequently the convergence problem is more serious for $p$-wave and 
$d$-wave scattering since a proportionally larger part of $Z_{\rm eff}$ 
comes from the $J \to \infty$ correction \cite{bromley03a,novikov04a}.  
We have not repeated the earlier $p$-wave calculations \cite{bromley03a} 
with a larger radial basis, but a comparison with the CI-Kohn data at 
$k = 0.4 \ a_0^{-1}$ indicates that the GL calculation again underestimates 
the impact  of the high $J$ orbitals.  The CI-Kohn calculation reported in      
\cite{bromley03a} gave $\Delta Z^{10}_{\rm eff} = 0.0480$ while the GL 
calculation gave 0.0416.  In addition, the CI-Kohn calculation gave 
$p = 1.852$ for $J = 10$ while the GL calculation gave $p = 2.152$ (it 
is likely that an infinite basis CI-Kohn calculation would have $p < 1.80$ 
at $J = 10$).  The GL calculation (which gave $Z_{rm eff} = 1.607$) 
underestimates the $L = 1$ $Z_{\rm eff}$ 
by about 0.18 at $k = 0.4$ $a_0^{-1}$ (the $T$-matrix calculation gives 
1.786 \cite{mitroy95a,ryzhikh00a} while van Reeth {\em et.al.} gave 1.794
\cite{vanreeth97a,vanreeth98a,gribakin04b}).  It is likely that at least 
0.10 of the discrepancy will arise from orbitals with $J \ge 10$.        

One of the major results of the GL calculations was their demonstration 
that the enhancement factor is independent of energy.  The enhancement 
factor can be defined as the factor that the annihilation rate, calculated 
as a simple product of the electron and positron densities needs to be 
increased in order to agree with the exact annihilation rate 
\cite{puska94,mitroy02a,novikov04a}.  They (GL) based this conclusion solely 
on a forensic analysis of the $Z_{\rm eff}$ annihilation rate matrix element.  
Figure 13 of GL reveals  that the variation of the $d$-wave enhancement 
factor with energy is noticeably larger than either the $s$-wave or $p$-wave 
enhancement factor \cite{gribakin04b}.  Since a larger fraction of the $d$-wave 
$Z_{\rm eff}$ comes from $J \ge 10$, the possibility exists that this stronger 
energy dependence is due to extrapolation issues as opposed to dynamical effects.
Although GL seem unaware of the result, the slow variation of the 
enhancement factor with energy had been demonstrated in a model potential 
analysis \cite{mitroy02a}.  Comparisons of model potential calculations 
with ab-initio variational and polarized orbital calculations had shown 
that a model potential calculation tuned to reproduce the energy dependence 
of the phase shifts also gave the energy dependence of $Z_{\rm eff}$ 
\cite{mitroy02a}.  The variation of the $d$-wave enhancement factor was 
determined by tuning a model potential to the large basis phase shifts of 
\cite{mitroy95b} and then normalizing to a similar calculation of $Z_{\rm eff}$ 
\cite{ryzhikh00a}.  The variation in the $d$-wave enhancement factor over the 
energy range from $k = 0$ to 0.5 $a_0^{-1}$ was less than 4$\%$.  Although 
the model potential result is not conclusive, it does appear that the 
variation in the $d$-wave $Z_{\rm eff}$ is less than that indicated by the 
GL calculation.  

Another area where application of the GL methodology could lead to 
larger than anticipated errors is in the determination the angular 
correlation or the $\gamma$ energy spectrum \cite{dunlop06a}.  These 
two properties depend 
on the relative momentum of the annihilating electron-positron pair 
\cite{charlton85,ryzhikh99a}.  It is known from investigations of 
momentum space wave functions that the low momentum part of the wave 
function largely arise from the large $r$ part of the wave function 
while the high momentum properties come from the small $r$ part of 
the wave function \cite{mccarthy91a}.  Under such circumstances, 
application of the GL method could easily result in errors to the 
$J \to \infty$ corrections that depend systematically on the 
$\gamma$-energy or recoil momentum.   

\section{Summary and Conclusions}

Single center methods represent a superficially attractive method 
to study mixed electron positron systems since existing computer
codes can be adapted without too much effort.  The penalty associated 
with this approach is the slow convergence of the binding energy 
and, more noticeably, the annihilation rate with respect to the 
partial wave expansion of the single particle basis.   
The results presented here are generally consistent with the 
asymptotic limits derived from second-order perturbation theory by 
Gribakin and Ludlow \cite{gribakin02a}.  The actual calculations 
at finite $J$ generally give increments to the energy (phase shift) 
and annihilation rate that decrease slightly slower than the GL
limits, but the overall trends are compatible with the GL limits.  

The tendency for the convergence with respect to the radial basis size to 
slow down as $J$ increases does have implications for the design of any CI 
type calculation.  Some sort of extrapolation in $J$ is necessary in order 
to determine the energy and more particularly the annihilation rate.   
But there is no point in making $J$ bigger if 
this is done at the expense of the radial basis set.  One simply
ends up with increments to the energy or annihilation rate 
which are systematically too small at higher $J$.  This problem 
does not seem to be restricted to the LTO basis used in the
present work.  Convergence problems at high $J$ are also present 
for the Gribakin and Ludlow calculations which used a $B$-spline 
basis \cite{gribakin04b} and the Saito calculations which 
used a natural orbital basis \cite{saito03a,saito05a,mitroy05i}.       
It is amusing to note that the one of the first manifestations of 
this problem occurred over 40 years ago \cite{tycko58a,schwartz62a}.  
       
The best methods for estimating the $J \to \infty$ corrections
depends on the quality of the underlying calculation.  For a
low precision calculation, Method I would seem to be appropriate.
A low precision calculation can probably be regarded as one with $p_E$ 
or $p_{\Gamma}$ exceeding 4 or 2 respectively when the $J \to \infty$ 
correction is evaluated (assuming that $p$ approaches its 
limiting value from below). Methods II or III would seem to be the 
preferred options for a high precision calculation.  As a general 
principle, inclusion of the second term in the asymptotic series 
leads to improved $\langle X \rangle^{\infty}$ predictions when 
compared with asymptotic series with the single term series.   
Method III is more susceptible to imperfections in the radial basis 
and should not be applied to the calculation of the annihilation
rate unless a very large radial basis set is employed.   
Irrespective of how the $J \to \infty$ corrections are evaluated, it 
is essential that the exponents $p$ relating the changes in the 
expectation values be examined as a test of the quality of the radial 
basis.     

The overall situation regarding the use of single center methods 
to compute positron-atom phase shifts or energies is that calculation 
to the sub 1$\%$ accuracy level is achievable for those systems that
have a parent atom ionization potential greater than 0.250 Hartree.  The
$O((J + {\scriptstyle \frac{1}{2}})^{-4})$ convergence means a $J$ 
of 10 or slightly larger will generally suffice as long as the method 
used to perform the $J \to \infty$ correction is more sophisticated 
than those used previously.   However, the situation with respect to 
the annihilation rate is much grimmer and it is not possible to
guarantee 1$\%$ accuracy for even the simple $e^+$-H system.  Here 
the $O((J + {\scriptstyle \frac{1}{2}})^{-2})$ convergence means 
the $J \to \infty$ correction is larger, and moreover the slow 
convergence with respect to the radial basis is further complicated 
by the fact that it is slower at high $J$ than low $J$.  
In this case, it appears that ``God is on the side of the big basis 
set'' \cite{boneparte}.

\begin{acknowledgments}

This work was supported by a research grant from the Australian 
Research Council.  The authors would like to thank Shane Caple 
for providing access to extra computing resources.  Dr Gribakin 
kindly provided tabulations of their calculations of $Z^J_{\rm eff}$.    

\end{acknowledgments}

%\bibliography{positron}

\begin{thebibliography}{56}
\bibitem[{bon()}]{boneparte}
\bibinfo{note}{This quote is based on a similar quote often attributed to N
  Boneparte}.
\end{thebibliography}

\end{document}